\documentclass[twocolumn]{aastex631}

\begin{document}
\turnoffeditone

\title[IMBHs in NGC 1399 GCs]{Prospects for Revealing Intermediate-Mass Black Holes in NGC 1399 using SKA}

\author{B. Karimi}
\affiliation{Canada Cambridge Academy, Markham, ON, Canada}

\author[0000-0003-2767-0090]{P. Barmby}
\affiliation{Department of Physics \& Astronomy, Western University, London, ON, Canada}
\affiliation{Institute for Earth and Space Exploration, Western University, London, ON, Canada}

\author[0000-0003-0428-2140]{S. Abbassi}
\affiliation{Department of Physics \& Astronomy, Western University, London, ON, Canada}

\correspondingauthor{P. Barmby}
\email{pbarmby@uwo.ca}

\begin{abstract}
This study investigates the detectability of intermediate-mass black holes (IMBHs) within the mass range $10^2 M_\odot \leq M_{\rm BH} \leq 10^5 M_\odot$ in the globular star clusters of NGC 1399 at a frequency of 300.00 MHz. Employing the theoretical Bondi accretion model and the empirical fundamental plane of black hole accretion, we estimate IMBH masses based on bolometric luminosity and X-ray/radio luminosities, respectively. By simulating a 3-hour observation of 77 globular cluster candidates using the Square Kilometer Array, we identify radio detection benchmarks indicative of accretion onto IMBHs. Our results show that IMBHs inside the globular star clusters located in NGC 1399 are indeed detectable, with the Bondi accretion model providing IMBH mass estimates ranging from $2.93 \times 10^{3.0 \pm 0.39} M_\odot$ to $7.43 \times 10^{4.0 \pm 0.39} M_\odot$, and the empirical fundamental-plane relation suggesting IMBH mass estimation with $3.41\times 10^{5.0 \pm 0.96} M_\odot$. These findings highlight the presence and detectability of IMBHs in globular clusters, offering insights into their role as precursors to supermassive black holes and enriching our understanding of black hole formation and evolution in astrophysical environments.
\end{abstract}


\section{Introduction}
Observational signatures facilitate the widespread identification of two distinct categories of astrophysical black holes: stellar-mass black holes (SBHs) and supermassive black holes (SMBHs) within the Universe. 
Stellar-mass black holes, with masses up to $10^2 M_\odot$, are observed within the Milky Way. 
Supermassive black holes, with masses exceeding $10^6 M_\odot$, reside at the cores of galaxies \citep[for reviews see][]{2001tysc.confE..87M, 2011NewAR..55..166F, 2020ARA&A..58..257G}.
Intermediate-Mass Black Holes (IMBHs) are defined as having masses  $10^2 M_\odot \leq M_{\rm IMBH} \leq 10^5 M_\odot$ \citep{2002MNRAS.330..232C, 2004ApJ...604..632G, 2021ApJ...918...18W}. 
The detection and characterization of IMBHs would provide a crucial link between the extensively studied SBHs and SMBHs \citep{2004A&A...414..895F}.  
If IMBHs indeed exist, they could potentially serve as seeds for the formation of supermassive black holes and the development of galactic bulges \citep{koliopanos2018intermediate}. 
Scenarios proposed for the formation of IMBHs include
the direct collapse of gas in atomic cooling halos in the early universe \citep[see, e.g.,][]{2006MNRAS.370..289B, 2014MNRAS.443.2410F}, the collapse of extremely massive Population III stars \citep{2001ApJ...551L..27M}, and the interaction of stars in the center of dense star clusters \citep{2002ApJ...576..899P,2002MNRAS.330..232C,2004MNRAS.351.1049M,2015MNRAS.454L..26H, 2024ApJ...969...29G}.

Observational attempts to identify IMBHs typically concentrate on astrophysical environments including ultra-/hyper-luminous X-ray sources, globular clusters (GCs), and dwarf galaxies
\citep[][]{2021MNRAS.504.1545D,2022NatAs...6...26R,Ward_2022,2023MNRAS.518.3386T}.
Dwarf galaxies, having undergone minimal mergers and evolving largely in isolation, harbor central black holes in nascent stages, potentially elucidating the evolutionary pathways towards supermassive black holes from IMBH seeds \citep{2004MNRAS.351.1049M,2008AIPC.1010..348M,2020ARA&A..58...27I}. 
GCs emerge as particularly promising sites for IMBH exploration, having long been recognized as hosts of X-ray binary (XRB) systems \citep{1975ApJ...199L.143C,2013AJ....146..135A}, and
offering critical insights into the formation mechanisms of massive black holes during the early universe \citep[for comprehensive reviews, see][]{2001ApJ...562L..19E,2020ARA&A..58..257G}. 
Debates persist regarding the presence of IMBHs in GCs: for example, \citet{2008ApJ...686..829H} and \citet{2018ApJ...867..119F} suggested that most GCs are unable to retain IMBHs against gravitational wave or Newtonian recoils from mergers; to explore this issue more deeply, it is necessary to study a large number of GCs per galaxy.
This underscores the ongoing importance of resolving the question of IMBH existence within astrophysical contexts \citep[e.g.,][]{2013MNRAS.430.2789P}.  

The X-ray and radio emission emanating from the central engines of stellar mass accreting X-ray binary systems are predominantly attributed to the presence of accretion disks (including coronae) and collimated jets. These emissions present a valuable avenue for investigating the correlation between the disk-jet dynamics and the mass of the black hole.  
Radio emission from  XRBs within GCs --- attributed to  distinctive non-thermal synchrotron radiation associated with jets --- has been successfully detected  by \citet{2024ApJ...961...54P}.
Recent observational efforts have focused on detecting IMBHs in GCs through various methods, including dynamical studies, X-ray emissions, and radio observations. For instance, 
\citet{2010ApJ...712L...1I} employed stellar dynamics to infer the presence of IMBHs in an NGC 1399 GC, while \citet{2024ApJ...961...54P} highlighted the importance of multi-wavelength observations in identifying and confirming IMBH candidates.
\citet{2013ApJ...776..118S} proposed that current radio telescopes can not detect radio signals from IMBHs in GCs.
\citet{2021ApJ...918...18W} performed a simulation study on Next Generation Very Large Array (ngVLA) detectability of IMBHs in GCs around NGC~4472 and concluded that individual $\sim 10^5M_\odot$ IMBHs would be detectable, with detections to  $\sim10^{4.5}M_\odot$ possible through radio stacking. 
Massive elliptical galaxies like NGC~4472 are particularly promising environments for such studies due to their rich GC systems and higher likelihood of hosting IMBHs \citep[see, e.g.,][]{2013ApJ...772...82H,2021ApJ...918...18W}.

The aim of this work is to determine whether IMBHs are detectable in NGC 1399 GCs in future observations with the SKA. 
We chose NGC 1399 due to its rich GC system, its location in the Southern Hemisphere, and the previous detection of a GC black hole \citep{2010ApJ...721..323S}. These factors make NGC 1399 an excellent candidate for searching for possible IMBHs using current and future radio telescopes. 
The SKA, with its unprecedented sensitivity and resolution, is well-suited for detecting the faint radio emissions that may be associated with accretion processes onto IMBHs. By targeting NGC 1399, a massive elliptical galaxy with a rich GC system, we aim to improve the statistical constraints on the presence and properties of IMBHs in GCs, thereby enhancing our understanding of their formation and retention mechanisms.
In this study, we employed two different methods: the theoretical Bondi accretion model \citep{2003ApJ...587L..35H, 2013ApJ...776..118S, 2018ApJ...862...16T, 2018ASPC..517..743W} and the empirical fundamental-plane relation for the hard X-ray state, along with measurements of X-ray luminosities \citep{2003MNRAS.345.1057M, 2004MNRAS.351.1049M, 2009ApJ...706..404G, 2013ApJ...776..118S, 2019ApJ...871...80G, 2021ApJ...918...18W} to estimate radio luminosity and black hole mass. If the estimated black hole mass falls within the range of $10^2 M_\odot \le M_{BH} \le 10^5 M_\odot$, then the black hole is considered an IMBH.

This paper is organized as follows:
In Section~\ref{sec:data}, we describe NGC 1399 and the relevant X-ray observations of its GC system. In Section~\ref{sec:Methodology}, we detail the methodologies employed, including the Bondi accretion model, the Fundamental Plane formula, and a simulated observation of NGC 1399 at 300~MHz with the SKA. In Section~\ref{sec:Qur}, we quantify black hole masses using the two different methods and determine the radio luminosity detection thresholds, explaining them as signatures of accretion onto central IMBHs in globular clusters. Finally, in Section~\ref{sec:Discussion and conlusion} we provide our discussion and conclusions.

\section{Globular clusters in NGC 1399}
\label{sec:data}

NGC 1399 is a cD galaxy within the Fornax cluster at a distance of $20.68\pm 0.50$ Mpc and mass of $\sim$ $10^{10.89 \pm 0.01}M_\odot$ \citep{2020ApJS..248...31L}. 
It hosts a substantial population of 6000--6500 globular clusters \citep{1998MNRAS.293..325F, 2001ApJ...557L..35A,2007ApJ...662..525K,2011ApJ...736...90P},
with an effective GC system diameter of a few tens of kiloparsecs \citep{2006ARA&A..44..193B}.
As with most large galaxies, NGC 1399's GCs can be divided into red, metal-rich, and blue, metal-poor subpopulations \citep{2004AJ....127.2094R}.
The typical GC half-starlight diameter of $\sim 5$ pc corresponds to $\sim 50$ mas at NGC 1399's distance.

\textit{Chandra} X-ray observations \citep{2011ApJ...736...90P,Dago2014,2020ApJS..248...31L} offer invaluable insights into the X-ray properties of GCs in NGC 1399 and the formation characteristics of low-mass X-ray binaries (LMXBs) within them. A significant proportion of the 2--10 keV X-ray emission emanates from the central region of NGC 1399, covering an area of approximately $8'\times 8'$, corresponding to $48.17 \times 48.17$~kpc. A considerable fraction of this emission is attributed to LMXBs both in the galaxy field and within GCs \citep{2001ApJ...557L..35A, 2022A&A...664A..41R}. LMXBs exhibit an association with both blue and red globular clusters within NGC 1399, with the red clusters hosting between $60-70\%$ of all LMXBs \citep{2011ApJ...736...90P}. Some $16\%$ of red clusters and $5\%$ of blue clusters are associated with LMXBs \citep{Dago2014}. 
The large number of GC X-ray sources in NGC 1399 makes this galaxy a great candidate to search for radio emission from IMBHs. 
In particular, the brightest color-confirmed GC X-ray source in NGC~1399 has an X-ray luminosity exceeding $4\times 10^{39}$~erg~s$^{-1}$ \citep{2011ApJ...736...90P}. This source is situated within one of the most metal-rich GCs and exhibits consistent X-ray luminosity, suggesting the potential presence of multiple LMXBs within a single GC and providing an important impetus for deep radio observations.

\begin{figure}
\centering
\includegraphics[width=0.5\textwidth]{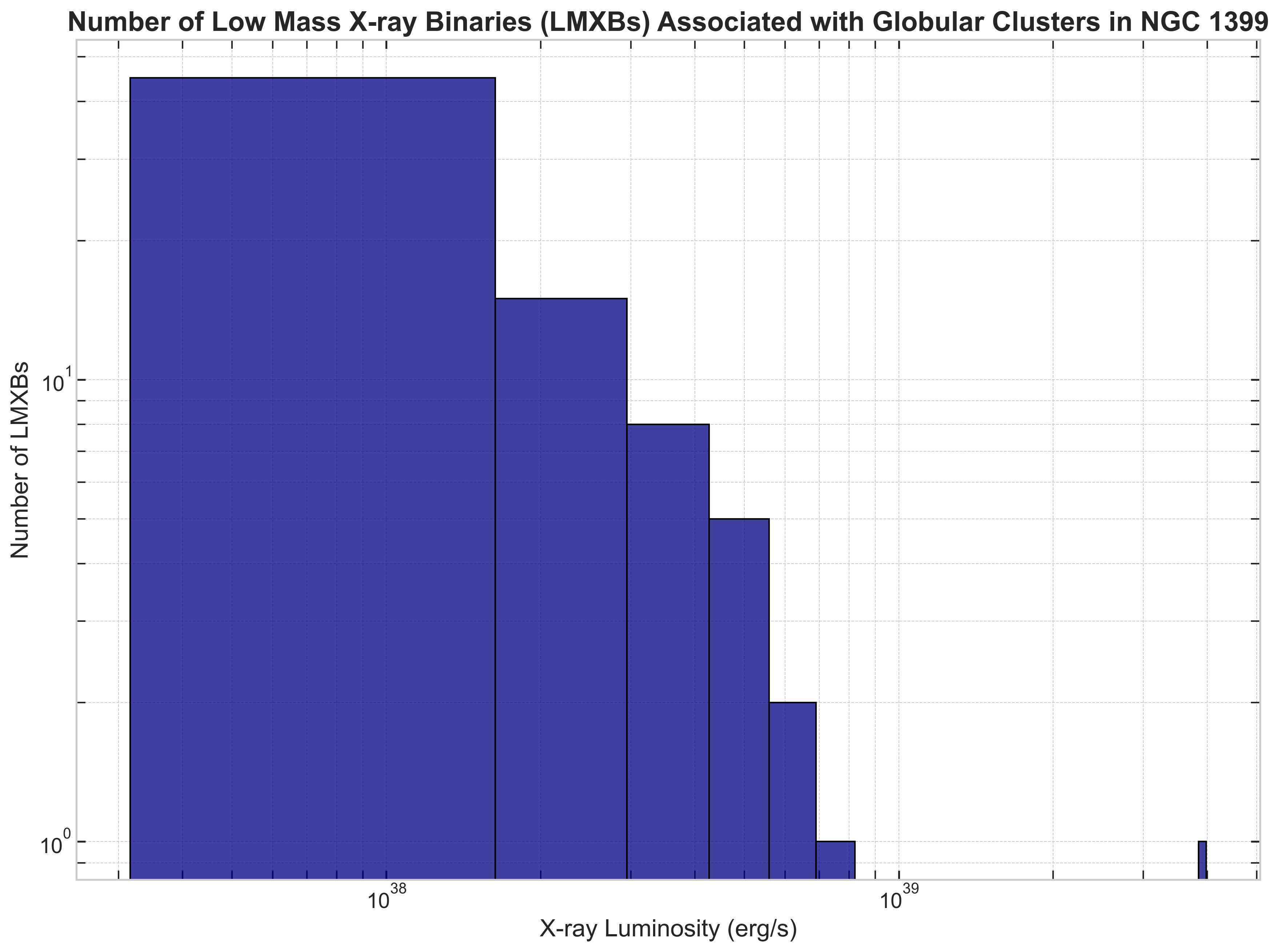}
\caption{Histogram showing the number of low-mass X-ray binaries  associated with blue and red globular clusters in NGC 1399 as a function of their X-ray luminosity \citep[data from][]{2020ApJS..248...31L}. The x-axis represents the X-ray luminosity in erg~s$^{-1}$, while the y-axis indicates the number of LMXBs. The distribution illustrates that the majority of LMXBs have an X-ray luminosity in the range of $10^{38}$ to $10^{39}$~erg~s$^{-1}$, highlighting the luminosity characteristics of these binaries within the globular clusters of NGC 1399.}  
\label{fig:sub1}
\end{figure}

 For our analysis in this study, we utilize the combined data from Chandra's X-ray observations of NGC~1399, as detailed in \citet{2020ApJS..248...31L}. This dataset was selected because it represents the deepest available X-ray observation of NGC 1399 GCs,  covering the largest area. 
 Figure \ref{fig:sub1} illustrates the luminosity distribution of LMXBs associated with globular clusters in NGC 1399. The histogram reveals that the majority of LMXBs exhibit X-ray luminosities between $10^{38}$~erg~s$^{-1}$ and $10^{39}$~erg~s$^{-1}$, with a noticeable decline at higher luminosities. This pattern suggests that lower luminosity LMXBs are more common within the globular clusters of NGC 1399, while higher luminosity LMXBs are relatively rare. The presence of LMXBs with X-ray luminosities up to $10^{39}$~erg~s$^{-1}$ hints at the existence of more massive compact objects or highly efficient accretion processes in some clusters.

\section{Methodology}
\label{sec:Methodology}

In this study, we employ two methods to estimate black hole masses in NGC~1399 GCs: 
\begin{enumerate}
    \item \textbf{Bondi Accretion Model}: We describe the classic Bondi accretion model for black holes, referencing works such as \citet{2003ApJ...587L..35H, 2013ApJ...776..118S, 2018ApJ...862...16T, 2021ApJ...918...18W, cho2023bridging}.
    
    \item \textbf{Fundamental Plane Formula}: We demonstrate the application of the fundamental plane formula as a mass predictor. This relation elucidates the correlation between a black hole's mass and its X-ray and radio luminosities, supported by studies such as \citet{2003MNRAS.345.1057M, 2004A&A...414..895F, 2006ApJ...645..890W, 2009ApJ...706..404G, 2012MNRAS.419..267P, 2019ApJ...871...80G, 2022MNRAS.516.6123G}. 
\end{enumerate}
If an estimated black hole mass falls within $10^2 M_\odot \le M \le 10^5 M_\odot$, we classify it as an IMBH. 
After estimating the black hole masses, we calculate their radio luminosities using the methodology outlined in \citet{2003MNRAS.345.1057M}, and detectability using the simulated observations described in Section~\ref{sec:ska}. Each method is described in detail below.

\subsection{Simulated Observation}
\label{sec:ska}

To simulate a radio observation of NGC 1399 GCs, we used the SKA continuum sensitivity calculator \citep{2022PASA...39...15S}.
We assumed a three-hour observation with SKA1-LOW at a wavelength of 1 metre ($\nu=300$~MHz). This frequency ensures adequate resolution (122~mas, corresponding to about 12.2~pc at the distance of NGC~1399) to distinguish individual GCs.
The Gaussian full width half maximum beam size ranges from 6 to 300 arcsec, and the field of view is 120 arcmin \citep{2019arXiv191212699B}, corresponding to 720~kpc at 20.7~Mpc, sufficient to encompass hundreds of GC candidates in a single SKA pointing.
The calculator gives a limiting flux density of $S_\nu = 100.5$~$\mu$Jy. With $\sigma = 33.5$~$\mu$Jy~beam$^{-1}$ \citep{2019arXiv191212699B}, we have $S_\nu = 3\sigma$, converting to a radio luminosity of $L_R = 1.617 \times 10^{34}$~erg~s$^{-1}$, assuming a spectral index $\alpha = 0$
\citep[a reasonable assumption for radio-emitting black hole jets;][]{2004MNRAS.351.1049M}.

\subsection{Bondi Accretion Model}
For our analysis, we employ the classical Bondi accretion model to estimate the black hole mass-accretion rate and subsequently the black hole mass. The gas characteristics at the Bondi radius and the black hole mass are critical parameters in determining the Bondi accretion rate. According to this model, a black hole in a GC accumulates mass from the surrounding tenuous gas at the Bondi accretion rate, which is consistent with the synchrotron radio emission observed from black holes in GCs \citep[see e.g.][]{2004MNRAS.351.1049M,  2018ASPC..517..743W}.

In the ideal spherical Bondi model, a black hole accretes ideal gas distributed locally from the GC environment. For simplicity, we do not account for the impact of globular star cluster's mass on the black hole accretion rate as it has been discussed in more detail by \citet{2013MNRAS.430.2789P}. Additionally, we assume the black hole is non-spinning. Despite its simplicity, this framework allows us to effectively use the Bondi model for our analysis.

To estimate the black hole accretion rate, we utilize the classical Bondi accretion model. Hydrodynamical simulations of gas flows in primordial galaxies, as shown by \citet{2020ApJ...894..141I}, establish an adiabatic process. For a specified black hole mass, gas is accreted at 3\% of the Bondi rate for a gas density of 0.2~particles~cm$^{-3}$ \citep{2001ApJ...557L.105F, 2018MNRAS.481..627A} and a temperature of $10^4$~K \citep{1975ApJ...197..147S, 2003ApJ...587L..35H}. 
After determining the black hole accretion rate, we estimate the black hole mass.

\subsection{Black Hole Mass Estimation Using Bondi Accretion Model}

To estimate black hole mass using the Bondi accretion model, we employ the relation between mass accretion rate and black hole mass (Equation \ref{equ: Bon}). Unlike previous studies \citep[e.g.,][]{2013ApJ...776..118S, 2021ApJ...918...18W}, which used the Fundamental Plane formula to estimate black hole mass after describing the Bondi model, we integrate both methods in our approach.

To estimate IMBH mass-accretion rate ($\dot{M}$), we establish a relationship between bolometric luminosity ($L^B$) and $\dot{M}$, considering a linear relation at low accretion rates. For accretion rates below $2\%$ of the Eddington rate, we employ a radiative efficiency of $\epsilon=0.1$ \citep{2019MNRAS.484.1724B,2021ApJ...918...18W}.
The Bondi accretion rate depends on gas density, temperature, and black hole mass  \citep{2013MNRAS.430.2789P}. By applying Equation~\ref{equ: Bon}, derived from \citet{2003ApJ...587L..35H}, with gas density $n=0.2$~cm$^{-3}$ and temperature $T=10^4$~K as noted by \citet{2013ApJ...776..118S}, we calculate the black hole mass. The uncertainty in black hole mass estimation using the Bondi accretion model is $0.39$ dex \citep{2021ApJ...918...18W}.
\begin{equation}
 \dot M_{BH}= 3.2 \times 10^{14}\times \left(\frac{M_{BH}}{2\times 10^3M_\odot}\right)^2
 \left(\frac{n}{0.2}\right)\left(\frac{T}{10^4}\right)^{-3/2} {\mathrm{kg\, s}}^{-1}   
 \label{equ: Bon}
\end{equation} 
In this study, we do not consider the influence of the magnetic field on black hole accretion rates as pointed out by \citet{cho2023bridging}, and solve the Bondi model for non-spinning black holes. Our analysis steps are listed below.

\begin{enumerate}
    \item We use the equation $L^B = \epsilon \dot{M} c^2$, where $L^B$ is the bolometric luminosity, $\epsilon = 0.1$ is the radiative efficiency, $\dot{M}$ is the mass accretion rate, and $c$ is the speed of light.
    \item The bolometric luminosity is related to the X-ray luminosity by $L_X (1 \sim 10\,\text{keV}) = \epsilon L_{Bol}$, with $\epsilon \sim 0.1$ \citep{2013ApJ...776..118S}. 
    \item Determining $L^B$ allows us to solve for $\dot{M}$ using the equation from step (1).
    \item Using the determined $\dot{M}$, we estimate the black hole mass with the Bondi formula (Equation \ref{equ: Bon}), assuming $n = 0.2$~cm$^{-3}$ and $T = 10^4$~K \citep{2013ApJ...776..118S}.
    \item We then use Equation \ref{eqn:fundplane2} to estimate the radio luminosity $L_R$ of the black hole, where $L_X$ (2--10 keV) is the X-ray luminosity and $M_{BH}$ is the black hole mass in units of $M_\odot$.
    Our Bondi method thus depends on the fundamental plane relationship \citep{2003MNRAS.345.1057M, 2004A&A...414..895F, 2012MNRAS.419..267P} as the only current method to link radio and X-ray luminosity with black hole mass.
\end{enumerate}

The Bondi model predicts a mass range for NGC 1399 GC black holes from $2.93 \times 10^{3.0 \pm 0.39} M_\odot$ to $7.43 \times 10^{4.0 \pm 0.39} M_\odot$. Given a typical globular cluster mass of $\sim 10^5 M_\odot$, these results from the Bondi model suggest that the black hole masses are within the expected range for IMBHs. The resulting  radio luminosities range from $10^{33.41 \pm 0.88}$~erg~s$^{-1}$ to $10^{34.96 \pm 0.88}$~erg~s$^{-1}$, which is around the detection threshold of radio luminosity $L_R = 1.617 \times 10^{34}$~erg~s$^{-1}$. Thus, we predict that radio emission from such IMBHs should be detectable with SKA observations.

\subsection{Fundamental Plane of Black Hole Activity}
\label{sec:FP}
Detecting non-thermal synchrotron radio emission from gas accretion by a central black hole is another approach to identifying IMBHs \citep[see e.g.][]{2004MNRAS.351.1049M, 2021ApJ...918...18W}. 
To estimate radio luminosities of possible IMBHs, we utilize the fundamental plane (FP) formula, which correlates X-ray luminosity, radio luminosity, and black hole mass \citep[see e.g.][]{2003MNRAS.345.1057M, 2009ApJ...706..404G, 2013ApJ...776..118S, 2019ApJ...871...80G, 2022MNRAS.516.6123G}.

The fundamental plane of black hole activity is applicable for the hard X-ray state (0.5--10~keV), validating its use in estimating black hole masses \citep{2003MNRAS.343L..59H, 2013ApJ...776..118S, 2021ApJ...918...18W, 2024ApJ...961...54P}. This formula relates radio emission from jet power, X-ray emission from the accretion disk, and black hole mass \citep[see e.g.][]{2003MNRAS.345.1057M, 2004MNRAS.351.1049M, 2004NewAR..48.1399F, 2006ApJ...645..890W, 2006NewA...11..567M, 2008ApJ...688..826L, 2009ApJ...706..404G, 2012MNRAS.419..267P, 2012ApJ...755L...1M,  2016ApJ...818..185F, 2017ApJ...836..104X, 2018ApJ...860..134Q, 2018ApJ...862...16T, 2019ApJ...871...80G, 2024ApJ...961...54P}.
The typical $L_X$ range used is 1--10~keV \citep{2003MNRAS.345.1057M, 2012MNRAS.419..267P}, representing a fraction of the bolometric luminosity \citep[see][]{2013ApJ...776..118S}. 
While \citet{2023arXiv231214098G} argue that the fundamental plane formula overestimates black hole mass in active galactic nuclei (AGN) with hard X-ray luminosity $L_{14-195\,\text{keV}} \ge 10^{42}\,\text{erg}\,\text{s}^{-1}$, they do not address its validity for globular clusters. Previous studies support its use for estimating IMBH masses in GCs \citep{2013ApJ...776..118S, 2021ApJ...918...18W, 2024ApJ...961...54P}.

\subsection{Quantifying Black Hole Mass using the Fundamental Plane Formula}

The fundamental plane formula demonstrates the correlation between radio luminosity, X-ray luminosity, and black hole mass \citep[see e.g.][]{2003MNRAS.345.1057M, 2009ApJ...706..404G, 2019ApJ...871...80G}. One form of this relation, which explains the correlation between radio flux, X-ray luminosity, black hole mass, and source distance, is given by \citet{2004MNRAS.351.1049M}:
\begin{equation}
S_\nu = 10 \left(\frac{L_X}{3\times 10^{31}}\right)^{0.6} \left(\frac{M_{BH}}{10^2 M_\odot}\right)^{0.78} \left(\frac{d}{10\, {\rm kpc}}\right)^{-2} \mu{\rm Jy}
\label{eqn:fundplane}
\end{equation}
where $S_\nu$ is the radio flux density in $\mu$Jy, $L_X$ is the X-ray luminosity in erg s$^{-1}$, $M_{BH}$ is the black hole mass in solar masses, and $d$ is the source distance in kpc. 
\citet{2019ApJ...871...80G} gives the uncertainty in black hole mass estimation using this formula as 0.96 dex and notes that uncertainty propagation in the FP relationship is complex. The uncertainties reported may underestimate the true errors due to the linear nature of the FP parameters when expressed in log space. 
Using the X-ray luminosities of 77 LMXBs from \citet{2020ApJS..248...31L}, we solve Equation \ref{eqn:fundplane} to estimate black hole masses. It is important to note that this form of the fundamental plane formula is applicable at low frequency.

Another form of the fundamental plane relation relates black hole mass, X-ray luminosity, and radio luminosity \citep{2003MNRAS.345.1057M}:
\begin{equation}
\log L_R = 7.33 + 0.6 \log L_X + 0.78 \log M_{BH}
\label{eqn:fundplane2}
\end{equation}
In this formula, $L_R$ and $L_X$ are in erg s$^{-1}$, and $M_{BH}$ is in solar masses. The uncertainty of the estimated radio luminosity is 0.88 dex \citep[see e.g.][]{2009ApJ...706..404G}, which we consider when presenting radio luminosity values. To estimate the mass of black holes detectable at 300 MHz, we follow these steps:
\begin{enumerate}
\item Set the estimated radio flux density at 20.7~Mpc to $S_\nu = 100.5$~$\mu$Jy, as explained in Section~\ref{sec:ska}.
\item Use X-ray luminosities, $L_X$, of 77 LMXBs taken from \citet{2020ApJS..248...31L} to estimate black hole mass in solar masses using Equation \ref{eqn:fundplane}.
\item Calculate the black hole radio luminosity using Equation \ref{eqn:fundplane2}.
\end{enumerate}

 \begin{figure}
\centering
\includegraphics[width=0.5\textwidth]{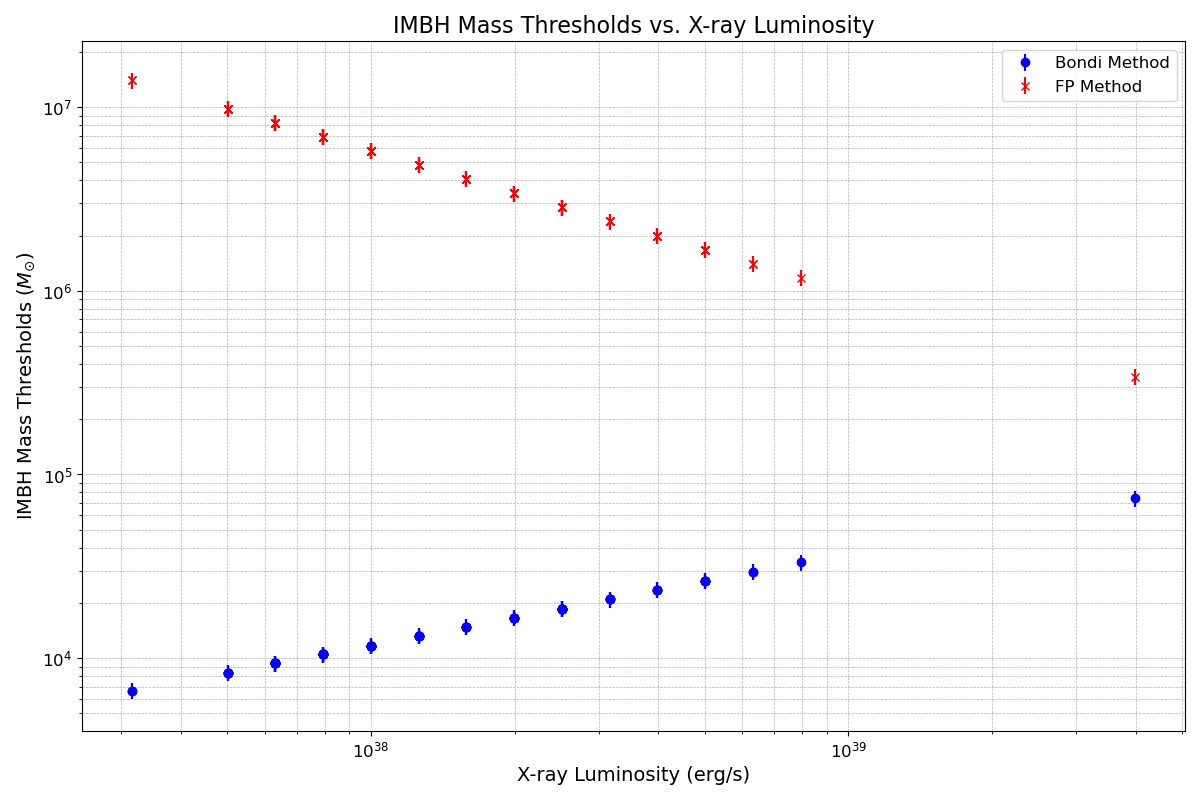}
\caption{Predicted black hole mass versus X-ray luminosity for intermediate-mass black holes  in NGC 1399, comparing estimates from the Bondi accretion model (blue circles) and the fundamental plane method (orange crosses). The Bondi model shows a positive correlation between mass and X-ray luminosity, while the FP method indicates higher masses with a broader range of X-ray luminosities, highlighting discrepancies between the two models in estimating black hole masses.}
\label{fig:sub2}
\end{figure}

\begin{figure}
\centering
\includegraphics[width=0.5\textwidth]{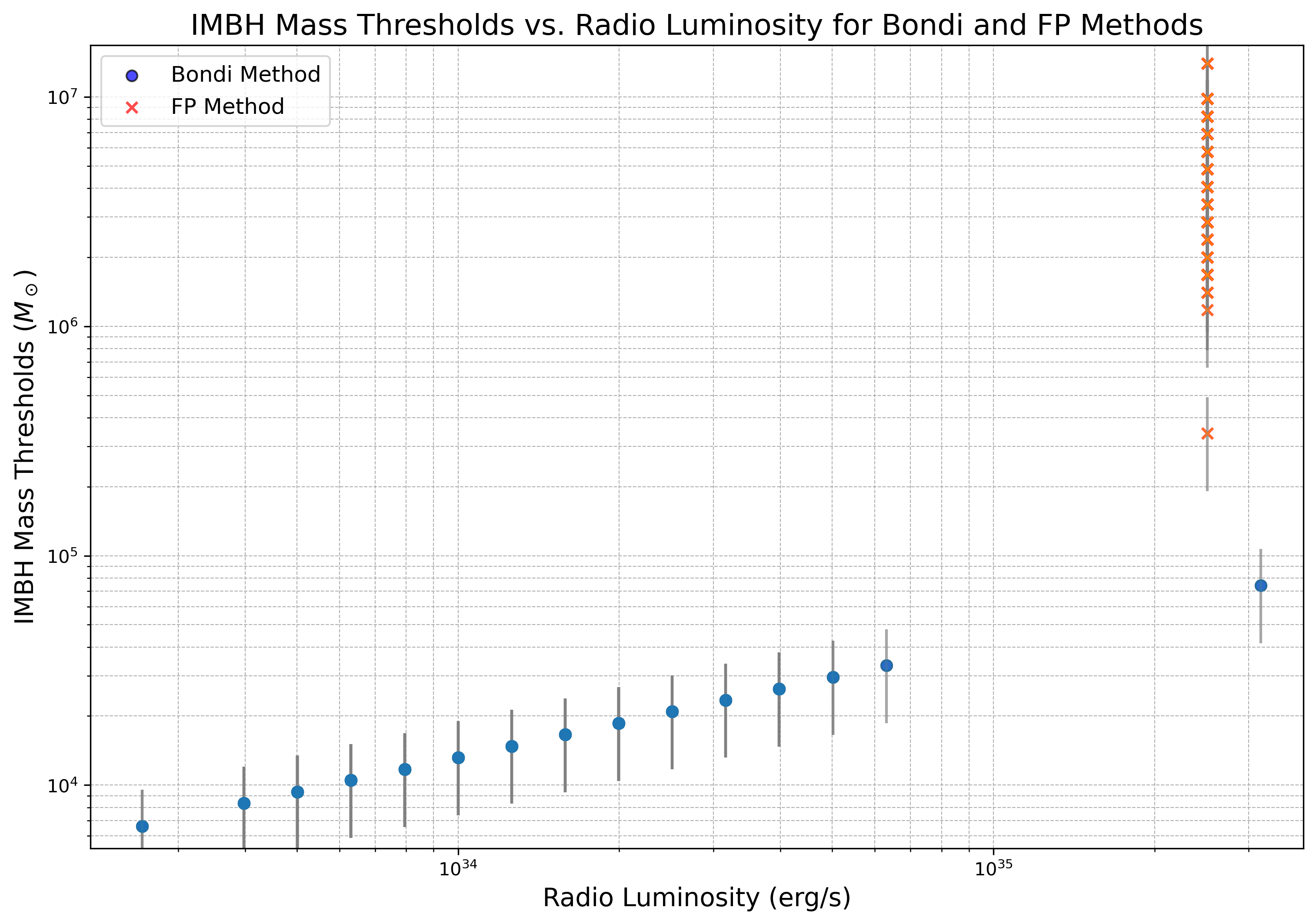}
\caption{Black hole mass versus radio luminosity for intermediate-mass black holes  estimated using the Bondi model and the Fundamental Plane formula. Black hole masses are given in terms of $M_\odot$. Radio luminosities are calculated using the Fundamental Plane relation and are given in erg s$^{-1}$. The minimum detectable radio luminosity is $L_R = 1.617 \times 10^{34}$ erg s$^{-1}$. }. 
\label{fig:sub3}
\end{figure}

\section{Results}
\label{sec:Qur}

Figure \ref{fig:sub2} depicts the relationship between black hole mass and X-ray luminosity for IMBHs in NGC 1399, using estimates from both the Bondi accretion model and the fundamental plane  method. The Bondi model demonstrates a clear positive correlation, indicating that higher mass IMBHs tend to have greater X-ray emissions. On the other hand, the FP method suggests higher masses across a broader range of X-ray luminosities, highlighting notable discrepancies between the two methods. These differences underscore the inherent uncertainties in the FP method, as noted by \citet{2019ApJ...871...80G}. Identifying LMXBs with luminosities $L_X \geq 10^{41}$ erg s$^{-1}$ as potential IMBH candidates further emphasizes the need for refined models to accurately characterize black hole properties.

In Figure \ref{fig:sub3}, we observe the relationship between black hole mass and radio luminosity for IMBHs, comparing the Bondi model (blue circles) and the FP method (orange crosses). The Bondi model indicates a positive correlation between mass and radio luminosity, while the FP method does not show a significant correlation. The Bondi model predicts 34 IMBHs are below the threshold of radio luminosity, making them undetectable, whereas the FP model predicts all 77 BHs are above the threshold, suggesting full detectability by SKA.

\begin{figure}
\centering
\includegraphics[width=0.5\textwidth]{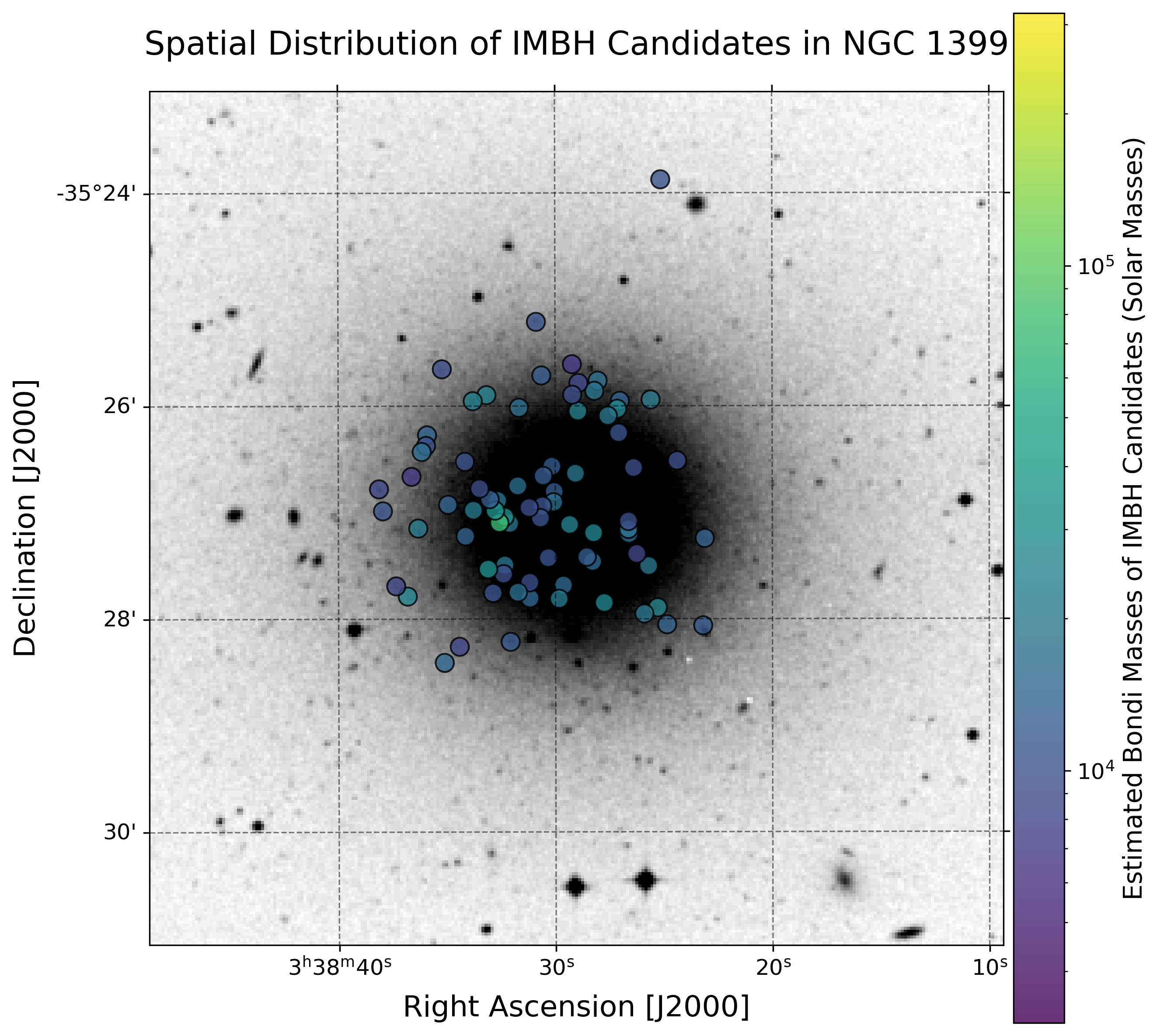}
\caption {Spatial distribution of predicted GC IMBHs in NGC 1399 overplotted on $8\arcmin\times 8\arcmin$ Digitized Sky Survey image. Sky positions of the GCs are taken from \citet{2020ApJS..248...31L}. The color gradation represents the estimated Bondi masses of the IMBHs in solar masses ($M_\odot$). 
}  
\label{fig:sub4}
\end{figure}

Figure \ref{fig:sub4} showcases the spatial distribution of predicted IMBH masses in NGC 1399, based on data from \citet{2020ApJS..248...31L}. The color gradation represents the estimated Bondi masses of these IMBHs, ranging from $10^3$ to $10^4$ $M_\odot$. This distribution indicates that IMBHs are concentrated in specific regions of the galaxy, likely within gravitational potential wells of globular clusters or dense stellar environments. The variation in masses points to diverse accretion histories and environments for these black holes. The fact that many IMBHs exceed the SKA detection threshold highlights the potential for future radio observations to deepen our understanding of their nature, formation, and role in galactic dynamics.

\begin{figure}
\centering
\includegraphics[width=0.5\textwidth]{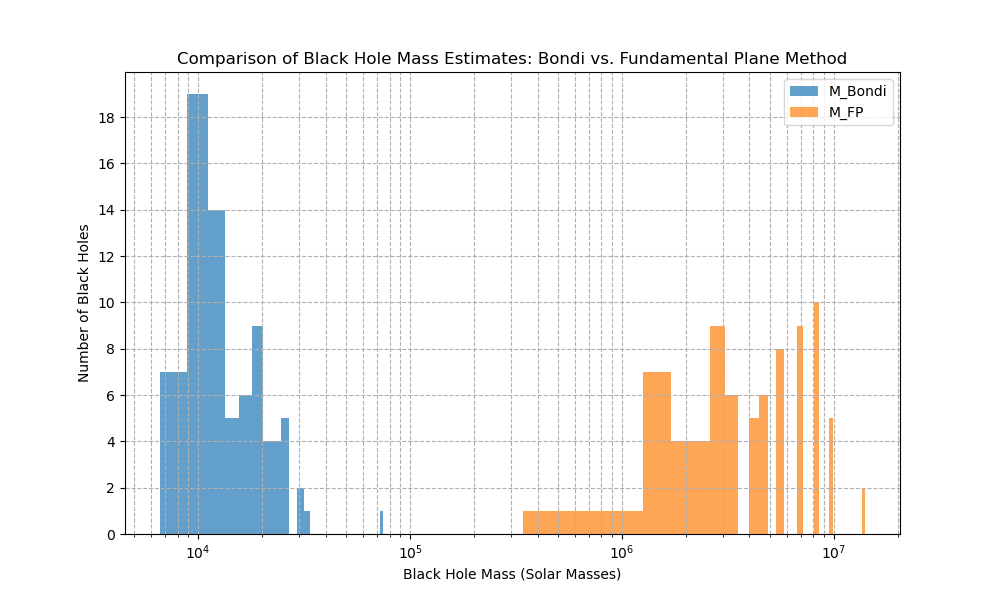}
\caption{Comparison of black hole mass estimates using the Bondi accretion model (blue) and the Fundamental Plane method (orange). The Bondi model predominantly estimates lower mass black holes ($10^3$ to $10^5$ solar masses), peaking around $10^4$ solar masses, whereas the FP method estimates higher masses ($10^6$ to $10^7$ solar masses). This highlights discrepancies between theoretical and empirical methods in black hole mass estimation.}. 
\label{fig:sub5}
\end{figure}

Finally, Figure \ref{fig:sub5} presents the distribution of IMBHs estimated using both the Bondi accretion model and the FP method. The Bondi model identifies all 77 low-mass X-ray binaries (LMXBs) as IMBHs, with masses ranging from $10^3$ to $10^4$ solar masses, suggesting a consistent presence of IMBHs across the sample. In contrast, the FP method identifies only one IMBH with mass of  $10^5$ solar masses, and 76 BHs with mass range of $10^{6-7}$ solar masses, highlighting a significant discrepancy between the two approaches. The Bondi model appears more inclusive, while the FP method sets a higher mass threshold. This underscores the importance of employing diverse methodologies in black hole studies to capture the full spectrum of IMBH detection and classification.

\section{Discussion and conclusions}
\label{sec:Discussion and conlusion}

Our work is consistent with that of \citet{2021ApJ...918...18W} in finding a gap between estimated IMBH  masses using the Bondi accretion model and the Fundamental Plane method.
This gap arises from inherent differences in their assumptions and methodologies. The Bondi model, relying on spherical accretion from hot gas with specific density and temperature, often underestimates the mass when real conditions deviate from this ideal. 
As noted by \citet{2021ApJ...918...18W}, gas densities in globular clusters are poorly known, with measurements existing for only two Milky Way GCs: 47 Tuc and M15.
Conversely, the FP method, which uses empirical correlations between black hole mass, X-ray luminosity, and radio luminosity, can overestimate the mass for high-luminosity sources due to its broad applicability assumptions \citep[see, e.g.,][]{2023arXiv231214098G}. These discrepancies underscore the need for multiple estimation methods to cross-verify black hole masses, accounting for the limitations and uncertainties of each approach.  

To address these limitations, recent advancements suggest modifying the Bondi model to include slow rotation, bridging the gap between classical Bondi and advection-dominated accretion flows (ADAFs). Studies have shown that incorporating slow rotation and external gravitational influences significantly reduces the accretion rate compared to the classical Bondi scenario \citep{2019MNRAS.489.3870S,2011MNRAS.415.3721N, 2022MNRAS.516.3984R,2023ApJ...954..117R}. This reduction could lower estimated black hole masses by one or two orders of magnitude, potentially widening the gap between the Bondi and FP model estimates \citep{2019MNRAS.484.1724B}. Observations of dense regions of massive elliptical galaxies, where hot gas rotates very slowly, support the existence of Bondi-type quasi-spherical accretion flows \citep{2017ApJ...834..148N, 2022ApJ...930L..12E}. This suggests similar processes could apply to IMBHs, leading to more accurate yet lower mass estimates with improved Bondi models. 

Therefore, incorporating realistic dynamics and external gravitational potentials into accretion models is crucial for refining our understanding of black hole masses across different regimes. Detailed numerical simulations are essential to validate these models and establish universal scaling relations in black hole accretion physics, ultimately enhancing our comprehension of IMBH characteristics and their role in galactic dynamics. To thoroughly address the discrepancies and refine mass estimates, it is imperative to employ multiple models and diverse samples, cross-verifying results to ensure robustness and accuracy in our understanding of black hole accretion processes.

Our study investigates the detectability of IMBHs within the GCs of NGC 1399 using the Square Kilometer Array (SKA). 
Our findings indicate that IMBHs within the GCs of NGC 1399 are detectable using relatively short observations with SKA-Low (flux density limit of 100.5 $\mu$Jy in a 3-hour observation), with significant implications for future observational campaigns. 
The SKA offers the possibility of both deeper and higher-frequency observations (with SKA-Mid); we defer detailed examination of these possibilities, and of observations of other systems such as ultra-compact dwarf galaxies \citep{2016hou}, to future work.
We employed two distinct methods for estimating IMBH masses: the theoretical Bondi accretion model and the empirical FP formula.
Our analysis using the Bondi model predicts IMBH masses in NGC 1399's GCs ranging from $2.93 \times 10^{3.0 \pm 0.39} M_\odot$ to $7.43 \times 10^{4.0 \pm 0.39} M_\odot$, while the FP formula suggests a broader range, with estimates up to $3.41 \times 10^{5.0 \pm 0.96} M_\odot$. The FP method also indicates higher radio luminosities, suggesting a greater likelihood of detecting these IMBHs with the SKA.

The significant discrepancies between the two methods are attributable to their underlying assumptions and methodologies.
The Bondi model, which assumes spherical accretion from hot gas with specific density and temperature, typically estimates lower black hole masses compared to the FP method. 
For higher X-ray luminosity, the gap between estimated mass from the Bondi model and the FP narrows to only one order of magnitude. 
The two methods are not entirely independent, since the Bondi model requires use of the fundamental plane to estimate radio luminosity; however
their discrepancies highlight the need for robust methods to accurately estimate black hole masses, accounting for the limitations and uncertainties inherent in each approach.
Recent advancements propose modifications to the Bondi model that include slow rotation and external gravitational influences. Studies have demonstrated that incorporating these factors significantly reduces the accretion rate compared to the classical Bondi scenario, potentially leading to even lower mass estimates for IMBHs. This underscores the importance of adopting more realistic models that account for the complexities of actual accretion processes.

The significant differences between the Bondi and FP model estimates highlight the need for developing more sophisticated models to improve the accuracy of black hole mass estimates. Detailed numerical simulations and multi-wavelength observations are essential to refine these models and establish robust scaling relations across different black hole mass regimes. 
By utilizing both theoretical and empirical models, our study offers valuable insights into the presence and properties of IMBHs in NGC 1399, contributing to the broader understanding of black hole formation and evolution in astrophysical environments. The SKA's unprecedented sensitivity and resolution will be instrumental in uncovering the faint radio emissions associated with accretion processes onto IMBHs, thereby improving statistical constraints on the presence and properties of IMBHs in GCs.

\begin{acknowledgments}
We thank the anonymous referee for a timely and helpful report.
\end{acknowledgments}

%

\vspace{5mm}
\facilities{CXO, HST}


\software{astropy \citep{2013A&A...558A..33A,2018AJ....156..123A}}

\bibliography{sample}{}

\begin{thebibliography}{}
\expandafter\ifx\csname natexlab\endcsname\relax\def\natexlab#1{#1}\fi
\providecommand{\url}[1]{\href{#1}{#1}}
\providecommand{\dodoi}[1]{doi:~\href{http://doi.org/#1}{\nolinkurl{#1}}}
\providecommand{\doeprint}[1]{\href{http://ascl.net/#1}{\nolinkurl{http://ascl.net/#1}}}
\providecommand{\doarXiv}[1]{\href{https://arxiv.org/abs/#1}{\nolinkurl{https://arxiv.org/abs/#1}}}

\bibitem[{{Abbate} {et~al.}(2018){Abbate}, {Possenti}, {Ridolfi}, {Freire}, {Camilo}, {Manchester}, \& {D'Amico}}]{2018MNRAS.481..627A}
{Abbate}, F., {Possenti}, A., {Ridolfi}, A., {et~al.} 2018, \mnras, 481, 627, \dodoi{10.1093/mnras/sty2298}

\bibitem[{{Agar} \& {Barmby}(2013)}]{2013AJ....146..135A}
{Agar}, J.~R.~R., \& {Barmby}, P. 2013, \aj, 146, 135, \dodoi{10.1088/0004-6256/146/5/135}

\bibitem[{{Angelini} {et~al.}(2001){Angelini}, {Loewenstein}, \& {Mushotzky}}]{2001ApJ...557L..35A}
{Angelini}, L., {Loewenstein}, M., \& {Mushotzky}, R.~F. 2001, \apjl, 557, L35, \dodoi{10.1086/323026}

\bibitem[{{Astropy Collaboration} {et~al.}(2013)}]{2013A&A...558A..33A}
{Astropy Collaboration}, {et~al.} 2013, \aap, 558, A33, \dodoi{10.1051/0004-6361/201322068}

\bibitem[{{Astropy Collaboration} {et~al.}(2018)}]{2018AJ....156..123A}
---. 2018, \aj, 156, 123, \dodoi{10.3847/1538-3881/aabc4f}

\bibitem[{{Begelman} {et~al.}(2006){Begelman}, {Volonteri}, \& {Rees}}]{2006MNRAS.370..289B}
{Begelman}, M.~C., {Volonteri}, M., \& {Rees}, M.~J. 2006, \mnras, 370, 289, \dodoi{10.1111/j.1365-2966.2006.10467.x}

\bibitem[{{Braun} {et~al.}(2019){Braun}, {Bonaldi}, {Bourke}, {Keane}, \& {Wagg}}]{2019arXiv191212699B}
{Braun}, R., {Bonaldi}, A., {Bourke}, T., {Keane}, E., \& {Wagg}, J. 2019, arXiv e-prints, arXiv:1912.12699, \dodoi{10.48550/arXiv.1912.12699}

\bibitem[{{Brodie} \& {Strader}(2006)}]{2006ARA&A..44..193B}
{Brodie}, J.~P., \& {Strader}, J. 2006, \araa, 44, 193, \dodoi{10.1146/annurev.astro.44.051905.09244110.48550/arXiv.astro-ph/0602601}

\bibitem[{{Bu} \& {Yang}(2019)}]{2019MNRAS.484.1724B}
{Bu}, D.-F., \& {Yang}, X.-H. 2019, \mnras, 484, 1724, \dodoi{10.1093/mnras/stz050}

\bibitem[{Cho {et~al.}(2023)Cho, Prather, Narayan, Natarajan, Su, Ricarte, \& Chatterjee}]{cho2023bridging}
Cho, H., Prather, B.~S., Narayan, R., {et~al.} 2023, Bridging Scales in Black Hole Accretion and Feedback: Magnetized Bondi Accretion in 3D GRMHD.
\newblock \doarXiv{2310.19135}

\bibitem[{{Clark}(1975)}]{1975ApJ...199L.143C}
{Clark}, G.~W. 1975, \apjl, 199, L143, \dodoi{10.1086/181869}

\bibitem[{{Dage} {et~al.}(2021){Dage}, {Kundu}, {Thygesen}, {Bahramian}, {Haggard}, {Irwin}, {Maccarone}, {Nair}, {Peacock}, {Strader}, \& {Zepf}}]{2021MNRAS.504.1545D}
{Dage}, K.~C., {Kundu}, A., {Thygesen}, E., {et~al.} 2021, \mnras, 504, 1545, \dodoi{10.1093/mnras/stab943}

\bibitem[{{D'Ago} {et~al.}(2014){D'Ago}, {Paolillo}, {Fabbiano}, {Puzia}, {Maccarone}, {Kundu}, {Goudfrooij}, \& {Zepf}}]{Dago2014}
{D'Ago}, G., {Paolillo}, M., {Fabbiano}, G., {et~al.} 2014, \aap, 567, A2, \dodoi{10.1051/0004-6361/201322722}

\bibitem[{{Ebisuzaki} {et~al.}(2001){Ebisuzaki}, {Makino}, {Tsuru}, {Funato}, {Portegies Zwart}, {Hut}, {McMillan}, {Matsushita}, {Matsumoto}, \& {Kawabe}}]{2001ApJ...562L..19E}
{Ebisuzaki}, T., {Makino}, J., {Tsuru}, T.~G., {et~al.} 2001, \apjl, 562, L19, \dodoi{10.1086/338118}

\bibitem[{{Event Horizon Telescope Collaboration} {et~al.}(2022){Event Horizon Telescope Collaboration}, {Akiyama}, {Alberdi}, {Alef}, {Algaba}, {Anantua}, {Asada}, {Azulay}, {Bach}, {Baczko}, {Ball}, {Balokovi{\'c}}, {Barrett}, {Baub{\"o}ck}, {Benson}, {Bintley}, {Blackburn}, {Blundell}, {Bouman}, {Bower}, {Boyce}, {Bremer}, {Brinkerink}, {Brissenden}, {Britzen}, {Broderick}, {Broguiere}, {Bronzwaer}, {Bustamante}, {Byun}, {Carlstrom}, {Ceccobello}, {Chael}, {Chan}, {Chatterjee}, {Chatterjee}, {Chen}, {Chen}, {Cheng}, {Cho}, {Christian}, {Conroy}, {Conway}, {Cordes}, {Crawford}, {Crew}, {Cruz-Osorio}, {Cui}, {Davelaar}, {De Laurentis}, {Deane}, {Dempsey}, {Desvignes}, {Dexter}, {Dhruv}, {Doeleman}, {Dougal}, {Dzib}, {Eatough}, {Emami}, {Falcke}, {Farah}, {Fish}, {Fomalont}, {Ford}, {Fraga-Encinas}, {Freeman}, {Friberg}, {Fromm}, {Fuentes}, {Galison}, {Gammie}, {Garc{\'\i}a}, {Gentaz}, {Georgiev}, {Goddi}, {Gold}, {G{\'o}mez-Ruiz}, {G{\'o}mez}, {Gu}, {Gurwell}, {Hada}, {Haggard}, {Haworth}, {Hecht}, {Hesper},
  {Heumann}, {Ho}, {Ho}, {Honma}, {Huang}, {Huang}, {Hughes}, {Ikeda}, {Impellizzeri}, {Inoue}, {Issaoun}, {James}, {Jannuzi}, {Janssen}, {Jeter}, {Jiang}, {Jim{\'e}nez-Rosales}, {Johnson}, {Jorstad}, {Joshi}, {Jung}, {Karami}, {Karuppusamy}, {Kawashima}, {Keating}, {Kettenis}, {Kim}, {Kim}, {Kim}, {Kim}, {Kino}, {Koay}, {Kocherlakota}, {Kofuji}, {Koch}, {Koyama}, {Kramer}, {Kramer}, {Krichbaum}, {Kuo}, {La Bella}, {Lauer}, {Lee}, {Lee}, {Leung}, {Levis}, {Li}, {Lico}, {Lindahl}, {Lindqvist}, {Lisakov}, {Liu}, {Liu}, {Liuzzo}, {Lo}, {Lobanov}, {Loinard}, {Lonsdale}, {Lu}, {Mao}, {Marchili}, {Markoff}, {Marrone}, {Marscher}, {Mart{\'\i}-Vidal}, {Matsushita}, {Matthews}, {Medeiros}, {Menten}, {Michalik}, {Mizuno}, {Mizuno}, {Moran}, {Moriyama}, {Moscibrodzka}, {M{\"u}ller}, {Mus}, {Musoke}, {Myserlis}, {Nadolski}, {Nagai}, {Nagar}, {Nakamura}, {Narayan}, {Narayanan}, {Natarajan}, {Nathanail}, {Fuentes}, {Neilsen}, {Neri}, {Ni}, {Noutsos}, {Nowak}, {Oh}, {Okino}, {Olivares}, {Ortiz-Le{\'o}n}, {Oyama},
  {{\"O}zel}, {Palumbo}, {Paraschos}, {Park}, {Parsons}, {Patel}, {Pen}, {Pesce}, {Pi{\'e}tu}, {Plambeck}, {PopStefanija}, {Porth}, {P{\"o}tzl}, {Prather}, {Preciado-L{\'o}pez}, {Psaltis}, {Pu}, {Ramakrishnan}, {Rao}, {Rawlings}, {Raymond}, {Rezzolla}, {Ricarte}, {Ripperda}, {Roelofs}, {Rogers}, {Ros}, {Romero-Ca{\~n}izales}, {Roshanineshat}, {Rottmann}, {Roy}, {Ruiz}, {Ruszczyk}, {Rygl}, {S{\'a}nchez}, {S{\'a}nchez-Arg{\"u}elles}, {S{\'a}nchez-Portal}, {Sasada}, {Satapathy}, {Savolainen}, {Schloerb}, {Schonfeld}, {Schuster}, {Shao}, {Shen}, {Small}, {Sohn}, {SooHoo}, {Souccar}, {Sun}, {Tazaki}, {Tetarenko}, {Tiede}, {Tilanus}, {Titus}, {Torne}, {Traianou}, {Trent}, {Trippe}, {Turk}, {van Bemmel}, {van Langevelde}, {van Rossum}, {Vos}, {Wagner}, {Ward-Thompson}, {Wardle}, {Weintroub}, {Wex}, {Wharton}, {Wielgus}, {Wiik}, {Witzel}, {Wondrak}, {Wong}, {Wu}, {Yamaguchi}, {Yoon}, {Young}, {Young}, {Younsi}, {Yuan}, {Yuan}, {Zensus}, {Zhang}, {Zhao}, {Zhao}, {Agurto}, {Allardi}, {Amestica}, {Araneda}, {Arriagada},
  {Berghuis}, {Bertarini}, {Berthold}, {Blanchard}, {Brown}, {C{\'a}rdenas}, {Cantzler}, {Caro}, {Castillo-Dom{\'\i}nguez}, {Chan}, {Chang}, {Chang}, {Chang}, {Chang}, {Chen}, {Chilson}, {Chuter}, {Ciechanowicz}, {Colin-Beltran}, {Coulson}, {Crowley}, {Degenaar}, {Dornbusch}, {Dur{\'a}n}, {Everett}, {Faber}, {Forster}, {Fuchs}, {Gale}, {Geertsema}, {Gonz{\'a}lez}, {Graham}, {Gueth}, {Halverson}, {Han}, {Han}, {Hasegawa}, {Hern{\'a}ndez-Rebollar}, {Herrera}, {Herrero-Illana}, {Heyminck}, {Hirota}, {Hoge}, {Hostler Schimpf}, {Howie}, {Huang}, {Jiang}, {Jinchi}, {John}, {Kimura}, {Klein}, {Kubo}, {Kuroda}, {Kwon}, {Lacasse}, {Laing}, {Leitch}, {Li}, {Liu}, {Liu}, {Lin}, {Lu}, {Mac-Auliffe}, {Martin-Cocher}, {Matulonis}, {Maute}, {Messias}, {Meyer-Zhao}, {Monta{\~n}a}, {Montenegro-Montes}, {Montgomerie}, {Moreno Nolasco}, {Muders}, {Nishioka}, {Norton}, {Nystrom}, {Ogawa}, {Olivares}, {Oshiro}, {P{\'e}rez-Beaupuits}, {Parra}, {Phillips}, {Poirier}, {Pradel}, {Qiu}, {Raffin}, {Rahlin}, {Ram{\'\i}rez}, {Ressler},
  {Reynolds}, {Rodr{\'\i}guez-Montoya}, {Saez-Madain}, {Santana}, {Shaw}, {Shirkey}, {Silva}, {Snow}, {Sousa}, {Sridharan}, {Stahm}, {Stark}, {Test}, {Torstensson}, {Venegas}, {Walther}, {Wei}, {White}, {Wieching}, {Wijnands}, {Wouterloot}, {Yu}, {Yu (于威)}, \& {Zeballos}}]{2022ApJ...930L..12E}
{Event Horizon Telescope Collaboration}, {Akiyama}, K., {Alberdi}, A., {et~al.} 2022, \apjl, 930, L12, \dodoi{10.3847/2041-8213/ac6674}

\bibitem[{{Falcke} {et~al.}(2004){Falcke}, {K{\"o}rding}, \& {Markoff}}]{2004A&A...414..895F}
{Falcke}, H., {K{\"o}rding}, E., \& {Markoff}, S. 2004, \aap, 414, 895, \dodoi{10.1051/0004-6361:20031683}

\bibitem[{{Fan} \& {Bai}(2016)}]{2016ApJ...818..185F}
{Fan}, X.-L., \& {Bai}, J.-M. 2016, \apj, 818, 185, \dodoi{10.3847/0004-637X/818/2/185}

\bibitem[{{Fender}(2004)}]{2004NewAR..48.1399F}
{Fender}, R. 2004, \nar, 48, 1399, \dodoi{10.1016/j.newar.2004.09.033}

\bibitem[{{Feng} \& {Soria}(2011)}]{2011NewAR..55..166F}
{Feng}, H., \& {Soria}, R. 2011, \nar, 55, 166, \dodoi{10.1016/j.newar.2011.08.002}

\bibitem[{{Ferrara} {et~al.}(2014){Ferrara}, {Salvadori}, {Yue}, \& {Schleicher}}]{2014MNRAS.443.2410F}
{Ferrara}, A., {Salvadori}, S., {Yue}, B., \& {Schleicher}, D. 2014, \mnras, 443, 2410, \dodoi{10.1093/mnras/stu1280}

\bibitem[{{Forbes} {et~al.}(1998){Forbes}, {Grillmair}, {Williger}, {Elson}, \& {Brodie}}]{1998MNRAS.293..325F}
{Forbes}, D.~A., {Grillmair}, C.~J., {Williger}, G.~M., {Elson}, R.~A.~W., \& {Brodie}, J.~P. 1998, \mnras, 293, 325, \dodoi{10.1046/j.1365-8711.1998.01202.x}

\bibitem[{{Fragione} {et~al.}(2018){Fragione}, {Leigh}, {Ginsburg}, \& {Kocsis}}]{2018ApJ...867..119F}
{Fragione}, G., {Leigh}, N. W.~C., {Ginsburg}, I., \& {Kocsis}, B. 2018, \apj, 867, 119, \dodoi{10.3847/1538-4357/aae486}

\bibitem[{{Freire} {et~al.}(2001){Freire}, {Kramer}, {Lyne}, {Camilo}, {Manchester}, \& {D'Amico}}]{2001ApJ...557L.105F}
{Freire}, P.~C., {Kramer}, M., {Lyne}, A.~G., {et~al.} 2001, \apjl, 557, L105, \dodoi{10.1086/323248}

\bibitem[{{Gliozzi} {et~al.}(2023){Gliozzi}, {Williams}, {Akylas}, {Papadakis}, {Shuvo}, {Halavatkar}, \& {Alt}}]{2023arXiv231214098G}
{Gliozzi}, M., {Williams}, J.~K., {Akylas}, A., {et~al.} 2023, arXiv e-prints, arXiv:2312.14098.
\newblock \doarXiv{2312.14098}

\bibitem[{{Gonz{\'a}lez Prieto} {et~al.}(2024){Gonz{\'a}lez Prieto}, {Weatherford}, {Fragione}, {Kremer}, \& {Rasio}}]{2024ApJ...969...29G}
{Gonz{\'a}lez Prieto}, E., {Weatherford}, N.~C., {Fragione}, G., {Kremer}, K., \& {Rasio}, F.~A. 2024, \apj, 969, 29, \dodoi{10.3847/1538-4357/ad43d6}

\bibitem[{{Greene} {et~al.}(2020){Greene}, {Strader}, \& {Ho}}]{2020ARA&A..58..257G}
{Greene}, J.~E., {Strader}, J., \& {Ho}, L.~C. 2020, \araa, 58, 257, \dodoi{10.1146/annurev-astro-032620-021835}

\bibitem[{{G{\"u}ltekin} {et~al.}(2009){G{\"u}ltekin}, {Cackett}, {Miller}, {Di Matteo}, {Markoff}, \& {Richstone}}]{2009ApJ...706..404G}
{G{\"u}ltekin}, K., {Cackett}, E.~M., {Miller}, J.~M., {et~al.} 2009, \apj, 706, 404, \dodoi{10.1088/0004-637X/706/1/404}

\bibitem[{{G{\"u}ltekin} {et~al.}(2019){G{\"u}ltekin}, {King}, {Cackett}, {Nyland}, {Miller}, {Di Matteo}, {Markoff}, \& {Rupen}}]{2019ApJ...871...80G}
{G{\"u}ltekin}, K., {King}, A.~L., {Cackett}, E.~M., {et~al.} 2019, \apj, 871, 80, \dodoi{10.3847/1538-4357/aaf6b9}

\bibitem[{{G{\"u}ltekin} {et~al.}(2022){G{\"u}ltekin}, {Nyland}, {Gray}, {Fehmer}, {Huang}, {Sparkman}, {Reines}, {Greene}, {Cackett}, \& {Baldassare}}]{2022MNRAS.516.6123G}
{G{\"u}ltekin}, K., {Nyland}, K., {Gray}, N., {et~al.} 2022, \mnras, 516, 6123, \dodoi{10.1093/mnras/stac2608}

\bibitem[{{G{\"u}rkan} {et~al.}(2004){G{\"u}rkan}, {Freitag}, \& {Rasio}}]{2004ApJ...604..632G}
{G{\"u}rkan}, M.~A., {Freitag}, M., \& {Rasio}, F.~A. 2004, \apj, 604, 632, \dodoi{10.1086/381968}

\bibitem[{{Harris} {et~al.}(2013){Harris}, {Harris}, \& {Alessi}}]{2013ApJ...772...82H}
{Harris}, W.~E., {Harris}, G. L.~H., \& {Alessi}, M. 2013, \apj, 772, 82, \dodoi{10.1088/0004-637X/772/2/82}

\bibitem[{{Heida} {et~al.}(2015){Heida}, {Jonker}, \& {Torres}}]{2015MNRAS.454L..26H}
{Heida}, M., {Jonker}, P.~G., \& {Torres}, M.~A.~P. 2015, \mnras, 454, L26, \dodoi{10.1093/mnrasl/slv121}

\bibitem[{{Heinz} \& {Sunyaev}(2003)}]{2003MNRAS.343L..59H}
{Heinz}, S., \& {Sunyaev}, R.~A. 2003, \mnras, 343, L59, \dodoi{10.1046/j.1365-8711.2003.06918.x}

\bibitem[{{Ho} {et~al.}(2003){Ho}, {Terashima}, \& {Okajima}}]{2003ApJ...587L..35H}
{Ho}, L.~C., {Terashima}, Y., \& {Okajima}, T. 2003, \apjl, 587, L35, \dodoi{10.1086/375042}

\bibitem[{{Holley-Bockelmann} {et~al.}(2008){Holley-Bockelmann}, {G{\"u}ltekin}, {Shoemaker}, \& {Yunes}}]{2008ApJ...686..829H}
{Holley-Bockelmann}, K., {G{\"u}ltekin}, K., {Shoemaker}, D., \& {Yunes}, N. 2008, \apj, 686, 829, \dodoi{10.1086/591218}

\bibitem[{{Hou} \& {Li}(2016)}]{2016hou}
{Hou}, M., \& {Li}, Z. 2016, \apj, 819, 164, \dodoi{10.3847/0004-637X/819/2/164}

\bibitem[{{Inayoshi} {et~al.}(2020{\natexlab{a}}){Inayoshi}, {Ichikawa}, \& {Ho}}]{2020ApJ...894..141I}
{Inayoshi}, K., {Ichikawa}, K., \& {Ho}, L.~C. 2020{\natexlab{a}}, \apj, 894, 141, \dodoi{10.3847/1538-4357/ab8569}

\bibitem[{{Inayoshi} {et~al.}(2020{\natexlab{b}}){Inayoshi}, {Visbal}, \& {Haiman}}]{2020ARA&A..58...27I}
{Inayoshi}, K., {Visbal}, E., \& {Haiman}, Z. 2020{\natexlab{b}}, \araa, 58, 27, \dodoi{10.1146/annurev-astro-120419-014455}

\bibitem[{{Irwin} {et~al.}(2010){Irwin}, {Brink}, {Bregman}, \& {Roberts}}]{2010ApJ...712L...1I}
{Irwin}, J.~A., {Brink}, T.~G., {Bregman}, J.~N., \& {Roberts}, T.~P. 2010, \apjl, 712, L1, \dodoi{10.1088/2041-8205/712/1/L1}

\bibitem[{Koliopanos(2018)}]{koliopanos2018intermediate}
Koliopanos, F. 2018, Intermediate Mass Black Holes: A brief review.
\newblock \doarXiv{1801.01095}

\bibitem[{{Kundu} {et~al.}(2007){Kundu}, {Maccarone}, \& {Zepf}}]{2007ApJ...662..525K}
{Kundu}, A., {Maccarone}, T.~J., \& {Zepf}, S.~E. 2007, \apj, 662, 525, \dodoi{10.1086/518021}

\bibitem[{{Lehmer} {et~al.}(2020){Lehmer}, {Ferrell}, {Doore}, {Eufrasio}, {Monson}, {Alexander}, {Basu-Zych}, {Brandt}, {Sivakoff}, {Tzanavaris}, {Yukita}, {Fragos}, \& {Ptak}}]{2020ApJS..248...31L}
{Lehmer}, B.~D., {Ferrell}, A.~P., {Doore}, K., {et~al.} 2020, \apjs, 248, 31, \dodoi{10.3847/1538-4365/ab9175}

\bibitem[{{Li} {et~al.}(2008){Li}, {Wu}, \& {Wang}}]{2008ApJ...688..826L}
{Li}, Z.-Y., {Wu}, X.-B., \& {Wang}, R. 2008, \apj, 688, 826, \dodoi{10.1086/592314}

\bibitem[{{Maccarone}(2004)}]{2004MNRAS.351.1049M}
{Maccarone}, T.~J. 2004, \mnras, 351, 1049, \dodoi{10.1111/j.1365-2966.2004.07859.x}

\bibitem[{{Maccarone} {et~al.}(2008){Maccarone}, {Kundu}, {Zepf}, {Shih}, {Rhode}, {Salzer}, \& {Bergond}}]{2008AIPC.1010..348M}
{Maccarone}, T.~J., {Kundu}, A., {Zepf}, S.~E., {et~al.} 2008, in American Institute of Physics Conference Series, Vol. 1010, A Population Explosion: The Nature \& Evolution of X-ray Binaries in Diverse Environments, ed. R.~M. {Bandyopadhyay}, S.~{Wachter}, D.~{Gelino}, \& C.~R. {Gelino}, 348--350, \dodoi{10.1063/1.2945073}

\bibitem[{{Madau} \& {Rees}(2001)}]{2001ApJ...551L..27M}
{Madau}, P., \& {Rees}, M.~J. 2001, \apjl, 551, L27, \dodoi{10.1086/319848}

\bibitem[{{Merloni} {et~al.}(2003){Merloni}, {Heinz}, \& {di Matteo}}]{2003MNRAS.345.1057M}
{Merloni}, A., {Heinz}, S., \& {di Matteo}, T. 2003, \mnras, 345, 1057, \dodoi{10.1046/j.1365-2966.2003.07017.x}

\bibitem[{{Merloni} {et~al.}(2006){Merloni}, {K{\"o}rding}, {Heinz}, {Markoff}, {Di Matteo}, \& {Falcke}}]{2006NewA...11..567M}
{Merloni}, A., {K{\"o}rding}, E., {Heinz}, S., {et~al.} 2006, \na, 11, 567, \dodoi{10.1016/j.newast.2006.03.002}

\bibitem[{{Miller} \& {Hamilton}(2001)}]{2001tysc.confE..87M}
{Miller}, M.~C., \& {Hamilton}, D.~P. 2001, in Two Years of Science with Chandra, ed. A.~{Siemiginowska}, 87

\bibitem[{{Miller} \& {Hamilton}(2002)}]{2002MNRAS.330..232C}
{Miller}, M.~C., \& {Hamilton}, D.~P. 2002, \mnras, 330, 232, \dodoi{10.1046/j.1365-8711.2002.05112.x}

\bibitem[{{Miller-Jones} {et~al.}(2012){Miller-Jones}, {Wrobel}, {Sivakoff}, {Heinke}, {Miller}, {Plotkin}, {Di Stefano}, {Greene}, {Ho}, {Joseph}, {Kong}, \& {Maccarone}}]{2012ApJ...755L...1M}
{Miller-Jones}, J.~C.~A., {Wrobel}, J.~M., {Sivakoff}, G.~R., {et~al.} 2012, \apjl, 755, L1, \dodoi{10.1088/2041-8205/755/1/L1}

\bibitem[{{Narayan} \& {Fabian}(2011)}]{2011MNRAS.415.3721N}
{Narayan}, R., \& {Fabian}, A.~C. 2011, \mnras, 415, 3721, \dodoi{10.1111/j.1365-2966.2011.18987.x}

\bibitem[{{Nielsen} {et~al.}(2017){Nielsen}, {Kacprzak}, {Muzahid}, {Churchill}, {Murphy}, \& {Charlton}}]{2017ApJ...834..148N}
{Nielsen}, N.~M., {Kacprzak}, G.~G., {Muzahid}, S., {et~al.} 2017, \apj, 834, 148, \dodoi{10.3847/1538-4357/834/2/148}

\bibitem[{{Paduano} {et~al.}(2024){Paduano}, {Bahramian}, {Miller-Jones}, {Kawka}, {Galvin}, {Rivera Sandoval}, {Kamann}, {Strader}, {Chomiuk}, {Heinke}, {Maccarone}, \& {Dreizler}}]{2024ApJ...961...54P}
{Paduano}, A., {Bahramian}, A., {Miller-Jones}, J. C.~A., {et~al.} 2024, \apj, 961, 54, \dodoi{10.3847/1538-4357/ad0e68}

\bibitem[{{Paolillo} {et~al.}(2011){Paolillo}, {Puzia}, {Goudfrooij}, {Zepf}, {Maccarone}, {Kundu}, {Fabbiano}, \& {Angelini}}]{2011ApJ...736...90P}
{Paolillo}, M., {Puzia}, T.~H., {Goudfrooij}, P., {et~al.} 2011, \apj, 736, 90, \dodoi{10.1088/0004-637X/736/2/90}

\bibitem[{{Pepe} \& {Pellizza}(2013)}]{2013MNRAS.430.2789P}
{Pepe}, C., \& {Pellizza}, L.~J. 2013, \mnras, 430, 2789, \dodoi{10.1093/mnras/stt080}

\bibitem[{{Plotkin} {et~al.}(2012){Plotkin}, {Markoff}, {Kelly}, {K{\"o}rding}, \& {Anderson}}]{2012MNRAS.419..267P}
{Plotkin}, R.~M., {Markoff}, S., {Kelly}, B.~C., {K{\"o}rding}, E., \& {Anderson}, S.~F. 2012, \mnras, 419, 267, \dodoi{10.1111/j.1365-2966.2011.19689.x}

\bibitem[{{Portegies Zwart} \& {McMillan}(2002)}]{2002ApJ...576..899P}
{Portegies Zwart}, S.~F., \& {McMillan}, S. L.~W. 2002, \apj, 576, 899, \dodoi{10.1086/341798}

\bibitem[{{Qian} {et~al.}(2018){Qian}, {Dong}, {Xie}, {Liu}, \& {Li}}]{2018ApJ...860..134Q}
{Qian}, L., {Dong}, X.-B., {Xie}, F.-G., {Liu}, W., \& {Li}, D. 2018, \apj, 860, 134, \dodoi{10.3847/1538-4357/aac32b}

\bibitem[{{Ranjbar} \& {Abbassi}(2023)}]{2023ApJ...954..117R}
{Ranjbar}, R., \& {Abbassi}, S. 2023, \apj, 954, 117, \dodoi{10.3847/1538-4357/ace163}

\bibitem[{{Ranjbar} {et~al.}(2022){Ranjbar}, {Mosallanezhad}, \& {Abbassi}}]{2022MNRAS.516.3984R}
{Ranjbar}, R., {Mosallanezhad}, A., \& {Abbassi}, S. 2022, \mnras, 516, 3984, \dodoi{10.1093/mnras/stac2454}

\bibitem[{{Reines}(2022)}]{2022NatAs...6...26R}
{Reines}, A.~E. 2022, Nature Astronomy, 6, 26, \dodoi{10.1038/s41550-021-01556-0}

\bibitem[{{Riccio} {et~al.}(2022){Riccio}, {Paolillo}, {Cantiello}, {D'Abrusco}, {Jin}, {Li}, {Puzia}, {Mieske}, {Prole}, {Iodice}, {D'Ago}, {Gatto}, \& {Spavone}}]{2022A&A...664A..41R}
{Riccio}, G., {Paolillo}, M., {Cantiello}, M., {et~al.} 2022, \aap, 664, A41, \dodoi{10.1051/0004-6361/202142894}

\bibitem[{{Richtler} {et~al.}(2004){Richtler}, {Dirsch}, {Gebhardt}, {Geisler}, {Hilker}, {Alonso}, {Forte}, {Grebel}, {Infante}, {Larsen}, {Minniti}, \& {Rejkuba}}]{2004AJ....127.2094R}
{Richtler}, T., {Dirsch}, B., {Gebhardt}, K., {et~al.} 2004, \aj, 127, 2094, \dodoi{10.1086/382721}

\bibitem[{{Samadi} {et~al.}(2019){Samadi}, {Zanganeh}, \& {Abbassi}}]{2019MNRAS.489.3870S}
{Samadi}, M., {Zanganeh}, S., \& {Abbassi}, S. 2019, \mnras, 489, 3870, \dodoi{10.1093/mnras/stz2397}

\bibitem[{{Scott} \& {Rose}(1975)}]{1975ApJ...197..147S}
{Scott}, E.~H., \& {Rose}, W.~K. 1975, \apj, 197, 147, \dodoi{10.1086/153496}

\bibitem[{{Shih} {et~al.}(2010){Shih}, {Kundu}, {Maccarone}, {Zepf}, \& {Joseph}}]{2010ApJ...721..323S}
{Shih}, I.~C., {Kundu}, A., {Maccarone}, T.~J., {Zepf}, S.~E., \& {Joseph}, T.~D. 2010, \apj, 721, 323, \dodoi{10.1088/0004-637X/721/1/323}

\bibitem[{{Sokolowski} {et~al.}(2022){Sokolowski}, {Tingay}, {Davidson}, {Wayth}, {Ung}, {Broderick}, {Juswardy}, {Kovaleva}, {Macario}, {Pupillo}, \& {Sutinjo}}]{2022PASA...39...15S}
{Sokolowski}, M., {Tingay}, S.~J., {Davidson}, D.~B., {et~al.} 2022, \pasa, 39, e015, \dodoi{10.1017/pasa.2021.63}

\bibitem[{{Sun} {et~al.}(2013){Sun}, {Jin}, {Gu}, {Liu}, {Lin}, \& {Lu}}]{2013ApJ...776..118S}
{Sun}, M.-Y., {Jin}, Y.-L., {Gu}, W.-M., {et~al.} 2013, \apj, 776, 118, \dodoi{10.1088/0004-637X/776/2/118}

\bibitem[{{Thygesen} {et~al.}(2023){Thygesen}, {Sun}, {Huang}, {Dage}, {Zepf}, {Kundu}, {Haggard}, \& {Maccarone}}]{2023MNRAS.518.3386T}
{Thygesen}, E., {Sun}, Y., {Huang}, J., {et~al.} 2023, \mnras, 518, 3386, \dodoi{10.1093/mnras/stac3244}

\bibitem[{{Tremou} {et~al.}(2018){Tremou}, {Strader}, {Chomiuk}, {Shishkovsky}, {Maccarone}, {Miller-Jones}, {Tudor}, {Heinke}, {Sivakoff}, {Seth}, \& {Noyola}}]{2018ApJ...862...16T}
{Tremou}, E., {Strader}, J., {Chomiuk}, L., {et~al.} 2018, \apj, 862, 16, \dodoi{10.3847/1538-4357/aac9b9}

\bibitem[{{Wang} {et~al.}(2006){Wang}, {Wu}, \& {Kong}}]{2006ApJ...645..890W}
{Wang}, R., {Wu}, X.-B., \& {Kong}, M.-Z. 2006, \apj, 645, 890, \dodoi{10.1086/504401}

\bibitem[{Ward {et~al.}(2022)Ward, Gezari, Nugent, Bellm, Dekany, Drake, Duev, Graham, Kasliwal, Kool, Masci, \& Riddle}]{Ward_2022}
Ward, C., Gezari, S., Nugent, P., {et~al.} 2022, \apj, 936, 104, \dodoi{10.3847/1538-4357/ac8666}

\bibitem[{{Wrobel} {et~al.}(2021){Wrobel}, {Maccarone}, {Miller-Jones}, \& {Nyland}}]{2021ApJ...918...18W}
{Wrobel}, J.~M., {Maccarone}, T.~J., {Miller-Jones}, J.~C.~A., \& {Nyland}, K.~E. 2021, \apj, 918, 18, \dodoi{10.3847/1538-4357/ac0ef3}

\bibitem[{{Wrobel} {et~al.}(2018){Wrobel}, {Miller-Jones}, {Nyland}, \& {Maccarone}}]{2018ASPC..517..743W}
{Wrobel}, J.~M., {Miller-Jones}, J.~C.~A., {Nyland}, K.~E., \& {Maccarone}, T.~J. 2018, in Astronomical Society of the Pacific Conference Series, Vol. 517, Science with a Next Generation Very Large Array, ed. E.~{Murphy}, 743

\bibitem[{{Xie} \& {Yuan}(2017)}]{2017ApJ...836..104X}
{Xie}, F.-G., \& {Yuan}, F. 2017, \apj, 836, 104, \dodoi{10.3847/1538-4357/aa5b90}

\end{thebibliography}
\bibliographystyle{aasjournal}


\end{document}